\documentstyle[prc,psfig,aps,twocolumn]{revtex}

\begin{document}

\title{
Nuclear fragmentation by tunneling
}

\author{
Toshiki Maruyama,$^{1,2}$ 
Aldo Bonasera,$^1$ 
and Satoshi Chiba$^2$
}

\address{
$^1$ 
Istituto Nazionale di Fisica Nucleare, Laboratorio Nazionale del Sud,
Via S.~Sofia 44, Catania 95123, Italy.\\
$^2$ 
Advanced Science Research Center, Japan Atomic Research Institute,
Shirakata Shirane 2-4, Ibaraki 319-1195, Japan.
}

\maketitle

\begin{abstract}
Fragmentation of nuclear system by tunneling is discussed
in a molecular dynamics simulation coupled with imaginary time method.
In this way we obtain informations on the fragmenting systems at low 
densities and temperatures.
These conditions cannot be reached  normally (i.e. above the barrier)
in nucleus-nucleus or nucleon-nucleus collisions.
The price to pay is the small probability of fragmentation by tunneling 
but we obtain observables which can be a clear signature of
such phenomena.
\end{abstract}

\bigskip

For a long time the problem of spontaneous fission has been of 
large interest for both experimentalists and theorists \cite{fission}.  
In fact spontaneous fission is a nice
example of a quantum many body problem.  
Both the $N$-particle nature of the system, the atomic nucleus,
and the quantum aspect, penetration of a barrier, 
has stimulated a huge literature production and
still the problem is largely unsolved from a quantum mechanical point of view.  
To a minor extent, the ``inverse process'' of SF i.e. subbarrier fusion 
has also stimulated many workers \cite{fission,kondrabonasera}.  
It is the purpose of this letter to suggest a similar problem 
which involves the quantum $N$-body features as
well as the possibility of critical phenomena.

Finite systems might show signatures which are typical of a phase transition.  
Experiments on fragmentation of atomic nuclei display typical features of 
a second order, power laws, critical exponents, etc..or a first order 
phase transition (caloric curve).
More recently similar features have been observed in 
metallic clusters and fullerens \cite{fulleren}.
These are systems that in the infinite size limit have an equation 
of state that resembles a Van Der Waals.  
Finite size effects smooth divergences which become maxima for some function. 
Nevertheless careful analysis are able to extract to a good accuracy the values 
of critical exponents which are very close to the liquid-gas values.  
The problem of these experiments is that the role of dynamics is dominant  
\cite{bruno}.
As a consequence it is not possible to prepare the system at all desired 
excitation energies, densities or temperatures as for a macroscopic system.  
For instance in heavy ion collisions if the energy of
 the particle is too low we observe fusion-fission of the two nuclei, 
while, if too large, fragmentation occurs.  
Imagine that in some way we have been able to prepare the 
system at density $\rho$ and temperature $T$.  
Because of the compression and/or thermal pressure, the system will expand.  
If the excitation energy is too low the expansion will come to 
an halt and the system will shrink back. 
This is some kind of monopole oscillation.   
On the other hand if the excitation is very large it will 
quickly expand and reach a region where the system is unstable 
and many fragments are formed.  
It is clear that in the expansion process the initial 
temperature will also decrease.  
We could roughly describe  this process with a collective coordinate $R(t)$, 
the radius of the system at time t and its conjugate coordinate.  
Here we are simply assuming that the expansion is spherical.  
These coordinates are somewhat the counterpart of the relative 
distance between fragments in the fission process \cite{kondrabonasera}. 
Similarly to the fission process we can imagine that connected to the 
collective variable $R(t)$ there is a collective potential $V(R)$.  
When the excitation energy is too low it means that 
we are below the maximum of the potential.  
That such a maximum exists comes, exactly as in the fission process,
from the short range nature of the nucleon nucleon force (or the typical 
Lenard-Jones force for metallic clusters) and the long range nature of Coulomb.  
Thus, similarly to SF, we can imagine to reach fragmentation 
by tunneling through the collective potential $V(R)$. 
When this happens, fragments will be formed at very low $T$ not reachable 
otherwise than through the tunneling effect.  
The price to pay as in all the subbarrier phenomena is a low cross section.  
We call the fragments formed via tunneling Quantum Drops (QD). 
Of course a necessary conditions to make such fragments is
to have enough energy and this could be easily estimated from the mass 
formula.  For instance one could imagine to search for the events where 3 
intermediate mass fragments (IMF) are formed (for which $Z\ge 3$) 
from an initial nucleus of mass $A$.  
>From these informations one can calculate the Q-value
of the reaction and the corresponding excitation energy to obtain IMF.  
The closer the excitation energy to the Q-value the smaller 
the probability to observe the events.  
It is the purpose of this paper to show through a microscopic simulation that 
this process is indeed possible and to discuss some features that 
undoubtedly characterize the formation of the Quantum Drops and 
that can be verified experimentally for instance in proton nucleus, 
gentle nucleus-nucleus or cluster-cluster collisions.

The microscopic simulational study using Vlasov equation
has been carried out for fusion and spontaneous fission 
problems \cite{kondrabonasera} below the Coulomb barrier.
The dynamics has been formulated introducing a one-dimensional 
collective degree of freedom which develops in real- and imaginary-time.
The method has reproduced well the fusion cross sections, fission kinetic
energies of the fission fragments and other observables \cite{kondrabonasera}.
The essential point is to introduce collective coordinates 
which enables us to treat the dynamics in imaginary time.

Since the expansion of the system can be parametrized in a one-dimensional
collective coordinate, we can, in principle, apply the imaginary-time 
method to its study.

In this paper we combine imaginary-time prescription with quantum 
molecular dynamics model to simulate fragmentation of finite 
system with relatively low excitation.


In QMD \cite{aich}, each particle follows the equations of motion 
given by the Hamiltonian.
The effective interaction in the Hamiltonian used in this paper is taken from
the QMD calculation of Ref.~\cite{Maru98} which can reproduce the 
saturation properties of nuclear matter, binding energies of ground state 
nuclei and the energy dependence of the optical potential.
To describe the sub-barrier dynamics of the system,
we treat the collective radial expansion in real- and imaginary-time
coupled to the motion of each particle.
If we do not allow the tunneling, the system will shrink back 
at the turning point if there is one, or the system continues to expand if
it has enough energy to overcome the attractive potential, 
i.e., there is no turning point.
What we do here is to simulate the tunneling motion when the 
collective variable reaches a turning point.

We treat initially spherical finite system.
Both the total center of mass coordinate and the total momentum are 0.
The collective coordinate $R_{\rm coll}$ and momentum $P_{\rm coll}$
for radial expansion is defined as
\begin{eqnarray}
R_{\rm coll} &\equiv& {1\over A}\sum_i\vec{\xi}_i\cdot{\bf R}_i,\\
P_{\rm coll} &\equiv& \sum_i \vec{\xi}_i\cdot{\bf P}_i,\\
\vec{\xi}_i &\equiv& {\bf R}_i/|R_i|.
\end{eqnarray}

The equation of motion for each particle is coupled to the 
collective variables as follows.
Before the collective radial motion reaches the turning point,
each particle obeys the normal equation of motion,
\begin{eqnarray}
\dot{{\bf R}_i} &=& \partial H/\partial{\bf P}_i,\\
\dot{{\bf P}_i} &=& -\partial H/\partial{\bf R}_i.
\end{eqnarray}
After reaching the turning point, 
the path of collective variables (real position and imaginary momentum)
is generally obtained by the inverse potential method.
We couple this inverse potential method for collective variables and
real-time evolution of each particle by introducing the collective
force $F_{\rm coll}$,
\begin{eqnarray}
\dot{{\bf R}_i} &=& \partial H/\partial{\bf P}_i,\\
\dot{{\bf P}_i} &=& -\partial H/\partial{\bf R}_i
-2F_{\rm coll} \vec{\xi}_i/A,\\
F_{\rm coll} &\equiv& \dot{P}_{\rm coll}.
\end{eqnarray}
Solving these equations of motion consistently, one obtains the real- 
and imaginary-time evolution of radial expansion.

When the collective expansion reaches the turning point again,
we turn off the inverse potential and solve real-time equation of motion.
If the collective variables never reach the turning point again,
the event is not accepted. Notice that if the system does not have 
enough energy to make fragments, then the second turning point 
will never be reached.

To obtain any physical quantity, tunneling events have to be counted
with their probability to happen, while events without tunneling can be
counted as to have probability 1.
The corresponding probability $\Pi$ of tunneling event is calculated
using the action $S$ during the tunneling process,
\begin{eqnarray}
\Pi &=& {1\over 1+e^{2S}}\\
S &=& \int\limits_{R_{\rm a}}^{R_{\rm b}} {P_{\rm coll} \over\hbar} 
dR_{\rm coll},
\end{eqnarray}
where $R_{\rm a}$ and $R_{\rm b}$ are starting and ending points of 
the tunneling region,
i.e., the two turning points.


\begin{figure}
\psfig{file=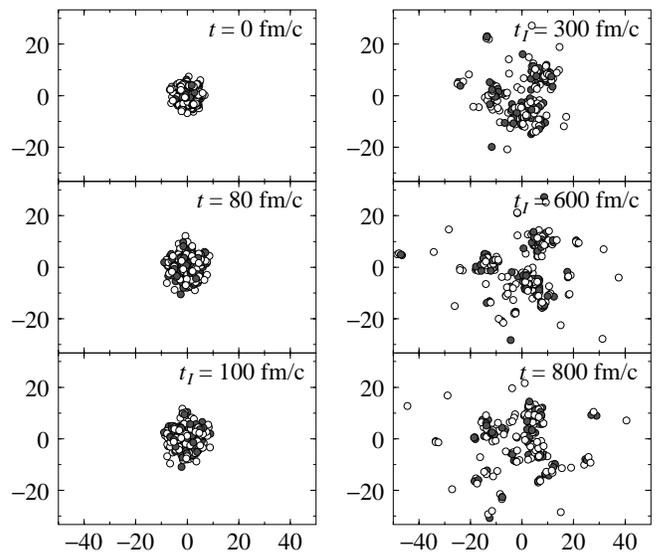,width=0.5\textwidth}
\caption{
Snapshot of tunneling fragmentation.
White circles represent neutron and gray ones protons.
The time in the simulation $t$ is shown in each panel.
After the system enters the tunneling region,
the time is written as $t_I$. 
}
\label{figPicture}
\end{figure}

We simulate the expansion of $^{230}\rm U$ system.
First we prepare the ground state of a nucleus and 
then compress uniformly to an excitation energy 
$E^*$ from 5 to 8 MeV/nucleon.
Due to the fluctuations between events caused by 
the different initial configurations of the nuclei, 
the potential energy during the expansion is 
different for different events.
Therefore the tunneling fragmentation occurs in some 
events where the potential energy is eventually high,
while there is no tunneling for events with lower potential energy.
At sufficiently low $E^*$, fragmentation occurs via tunneling only.
In Fig.~\ref{figPicture} we display snapshots of a typical tunneling event.  
The collective coordinate P(t) becomes zero at $t=100$ fm/c.  
At this stage we turn to imaginary times as described above and the
tunneling begins.  Notice that the system indeed expands and its shape 
can be rather well approximated to a sphere at the beginning.  
But already at 300 fm/c (in imaginary time) due to the molecular dynamics
nature of the simulation, the spherical approximation is broken.  
This shortcoming of our approach should be kept in mind because the 
calculated actions will be quite unrealistic.  
This is similar to the use of one collective coordinate 
(the relative distance between centers) in SF process.  
In that case many calculated features are quantitatively 
wrong but qualitatively acceptable.  
Since our Quantum Drops is a proposed novel mechanism we can only give
qualitative features and the model assumption must be refined when 
experimental data will start to be available.

\begin{figure}
\psfig{file=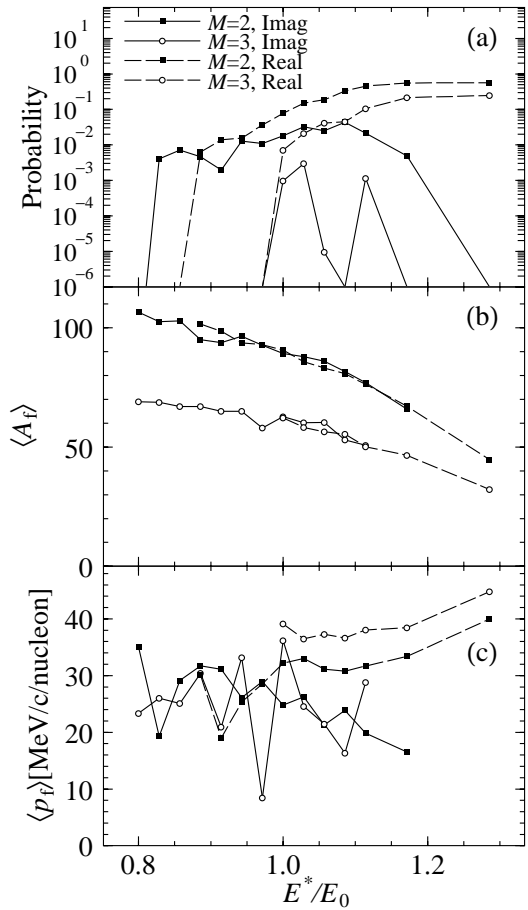,width=0.45\textwidth}
\caption{
(a) Reaction probability; (b) Average mass of large ($Z\geq3$) fragments;
(c) Average momenta of large fragments.
Note that the horizontal axis is shown by the normalized excitation
energy $E^*/E_0$, where $E_0=7.0$ MeV/nucleon (the threshold for $M=3$ events.)
}
\label{figMass}
\end{figure}
   
In fact the calculated action $S$ in the tunneling 
events for lower excitation energy is large and the 
probability of tunneling fragmentation is very small.
Figure \ref{figMass}(a) shows the calculated probability of 
fragmentation with fragment multiplicity of 2 and 3 vs.~ $E^*/E_0$.
Here $E_0$ is the threshold for having real events with $M=3$.
The value $E_0$ depends on the detail of the force and in our case $E_0=7$ MeV.
This seems to us a quite unrealistic value; in fact when  we compare our model to 
data on fragmentation (above the barrier) we underestimate the data, 
i.e.~the model is less explosive.
>From the data, for instance Aladin data \cite{aradin}, 
we can estimate $E_0\approx 3$ MeV/nucleon.

For the imaginary events the probabilities are obtained using 
the action $S$ of the tunneling process.
If only one type of reaction is considered the probability of
both the real and the imaginary events are 1/2 at the threshold (barrier) energy.
We see good agreement of probabilities for $M=2$ near the vanishing point
of real event.
But the value of the probability is not 1/2 since there are many types 
of events we have to consider and the barrier height is fluctuating.
For $M=3$ events this effect is serious and the probabilities for
real and imaginary events do not meet at any point.

\begin{figure}
\psfig{file=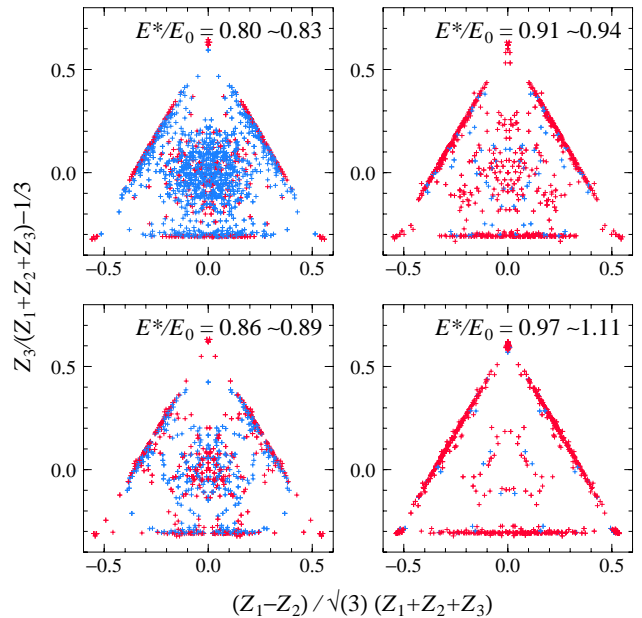,width=0.5\textwidth}
\caption{
(Color) Dalitz plot for fragment mass.
Red points indicate data from real fragmentation
and blue are imaginary process.
The excitation energy shown in each panel is normalized by 
threshold energy $E_0$.
}
\label{figDalitz}
\end{figure}

Although the probability of tunneling fragmentation is very small,
one possible observable may be the fragment mass distribution.
In Fig.~\ref{figMass}(b) we plot the average fragment mass for intermediate 
mass fragment multiplicity $M$ of 2 and 3.
Although we see slight difference between imaginary and real events,
the values are too close to distinguish.
Figure \ref{figDalitz} is a Dalitz plot of 
three largest fragment mass in each event.
Plotted with red color is fragmentation without 
tunneling process and blue by tunneling.
In this figure we can clearly distinguish both types of reactions.
When the fragmentation occurs by tunneling the size of fragments 
are more equally distributed (central region in Dalitz plot) 
than the case without tunneling process.
Equally distributed fragment size is not preferred by 
the Coulomb potential when position of fragments are 
concentrated in the early stage of fragmentation.
Such a situation is only possible with tunneling process.
Without tunneling process, on the other hand, the low-energy 
fragmentation is only possible 
through sequential decays.

Another possible observable to distinguish the tunneling events from
the real events is the momentum distribution of produced fragments 
shown in Fig.~\ref{figMass}(c).
Again we see significant difference between imaginary and real events.
It is rather natural that real events have larger value of fragment momenta 
since the eventual difference of kinetic energy of the expansion determines 
whether the system experience the imaginary process or not.
However, the difference of the process itself has some effects 
on the resulting fragment momenta:
In the real events the fragments momenta are larger for cases with 
large fragment multiplicity $M=3$ compared to $M=2$. 
This is due to the different Q-value for events with $M=3$ and 2.
In fact the difference of the total kinetic energy of large fragments
for two types of events, which amounts to about 30 MeV, almost exactly 
corresponds the difference of their Q-values estimated by a simple liquid-drop model.

In imaginary events, on the contrary, fragment momenta 
for $M=3$ and 2 are almost the same 
(within a numerical accuracy) though there are large fluctuations due
to the poor statistics.
This is due to the fact that below the threshold, the system must use all the
excitation energy to form the final fragments. 
In order to do that the final fragments keep the lowest
possible kinetic energy.
The final kinetic energy is essentially due to Coulomb repulsion at the
second turning point.
Thus precise experimental informations on the final kinetic
energies will give us an insight on the densities at freeze out.
The comparison of fragment momenta between the 
fragment multiplicity $M=2$ and 3 may give a signature of 
imaginary process in the fragmentation.

In conclusion, we have simulated the fragmentation process by tunneling.
The calculated probability of such phenomena is rather small,
and the threshold energy for fragmentation in our calculation 
has been found to be unrealistically high.
Therefore by employing more suitable interaction, the threshold energy
and even the tunneling probability could change.
Apart from the probability of the tunneling fragmentation,
we have proposed two possible observables for this phenomena.
One is the Dalitz plot of produced large fragments.
In tunneling fragmentation, fragment mass is more uniformly
distributed than that of normal fragmentation.
The other is the fragment momentum distribution,
where we have observed that the fragment momentum are
insensitive to the fragment multiplicity.


\begin{references}

\bibitem{fission}
{\it Proceedings of the International Conference on Fifty Years Research 
in Nuclear Fission, Berlin April 1989,}
edited by D.~Hilscher et al. [Nucl. Phys. {\bf A502}, 1c-639c (1989)].

\bibitem{kondrabonasera}
A.~Bonasera and V.~N.~Kondratyev, Phys. Lett. B. {\bf 339}, 207 (1994);
A.~Bonasera and A.~Iwamoto, Phys. Rev. Lett. {\bf 78}, 187 (1997).

\bibitem{fulleren}
A.~Bonasera, Physics World {\bf 12} No.~2, 20 (1999).

\bibitem{bruno}
A.~Bonasera, M.~Bruno, C.~O.~Dorso, and P.~F.~Mastinu, Rivista del Nuovo Cimento {\bf 23},
No.~2, p.~1 (2000).

\bibitem{aich}
J.~Aichelin, Phys. Rep. {\bf 202}, 233 (1991); and references therein.

\bibitem{Maru98}
T.~Maruyama, K.~Niita, K.~Oyamatsu, T.~Maruyama, S.~Chiba, and A.~Iwamoto,
Phys. Rev. C {\bf 57}, 655 (1998).

\bibitem{aradin}
J.~Pochodzalla et al., Phys. Rev. Lett. {\bf 75} 1040 (1995).

\end{references}
\end{document}